\documentstyle[floats,aps,epsf,prl]{revtex}

\begin{document}
\draft
\preprint{May 22, 1997}

\twocolumn[\hsize\textwidth\columnwidth\hsize\csname
@twocolumnfalse\endcsname

\title{Nonlocality of Kohn-Sham exchange-correlation fields in dielectrics}
\author{David Vanderbilt}
\address{Department of Physics and Astronomy, Rutgers University,
Piscataway, New Jersey 08855-0849}

\date{May 22, 1997}
\maketitle

\begin{abstract}
The theory of the macroscopic field appearing in the Kohn-Sham
exchange-correlation potential for dielectric materials, as
introduced by Gonze, Ghosez and Godby, is reexamined.  It is shown
that this Kohn-Sham field cannot be determined from a knowledge
of the local state of the material (local crystal potential, electric
field, and polarization) alone.  Instead, it has an intrinsically nonlocal
dependence on the global electrostatic configuration.  For example, it
vanishes in simple transverse configurations of a polarized dielectric,
but not in longitudinal ones.
\end{abstract}

\vskip1pc]

Ever since it became clear that electric polarization
is indeed a well-defined bulk quantity in an insulating crystal
\cite{ksv,vks,resta1,om}, the status of this electric polarization
in the Kohn-Sham (KS) density-functional theory (DFT) \cite{hk,ks}
has become a topic of considerable interest
\cite{vks,ggg1,ggg2,aulbur,resta2,mazin,mo}.  (Throughout this Letter,
I refer to the {\it exact} version of DFT in which the true
KS exchange-correlation (XC) functional is presumed known.)
In Ref.\ \cite{vks} it was argued that the polarization of a
crystalline insulator should be given exactly by DFT, on the basis
that charge densities are given exactly, and that any errors
in polarization would show up as errors in charge densities at
surfaces or interfaces.  Gonze, Ghosez and Godby (GGG)
\cite{ggg1,ggg2} then pointed out that in order for the DFT
polarization to be correct, the DFT XC potential would need to
have a linear spatial variation (i.e, an ``XC electric field''
$\cal E_{\rm xc}$ would have to be present).  They formulated a
new version of DFT appropriate for crystalline insulators, in
which the density $n({\bf r})$ {\it and} the electronic
polarization ${\bf P}$ are shown to be uniquely related to the
periodic part of the potential $\widetilde V({\bf r})$ {\it and} the
electric field $\cal E$.  This extended Hohenberg-Kohn (HK) \cite{hk}
principle then allows the XC energy to be expressed as a functional
of $n({\bf r})$ and ${\bf P}$, instead of just $n(\bf r)$
alone.  Recently, Martin and Ortiz \cite{mo} have reformulated and
extended this analysis.  While agreeing with many of
the conclusions of GGG, they nevertheless appear to express some
doubts about the GGG interpretation of the XC field, preferring
instead to focus on the HK and KS descriptions for the {\it change} in
polarization connected to a {\it change} in field.

In this Letter, I present an analysis that clarifies the role
of the XC field in the exact KS theory.
I start by deducing the behavior of the XC potential for
several simple configurations of a finite sample of spontaneously
polarized dielectric material in vacuum.  These examples illustrate
misleading aspects of certain arguments given by the previous
Refs.\ \cite{vks,ggg2,mo}.  Briefly, it is now understood
\cite{ggg1,ggg2,mo} that the local periodic charge density in some
small region of the sample can be generated by any of a continuous
family of KS potentials labeled by the choice of effective field
$\cal E_{\rm eff}=\cal E+E_{\rm xc}$ or, equivalently, by the choice
of electric polarization
${\bf P}_{\rm eff}$, in the same small region of the corresponding
fictitious KS system.  Based on Refs.\ \cite{vks} and \cite{ggg2},
one might assume that the correct choice would be the one that makes
the polarization correct, $\bf P_{\rm eff}=\bf P$; while
in a naive approach one would make the choice
$\cal E_{\rm eff}=E$, at least for the case ${\cal E}=0$.  Here,
I show that {\it neither of the above choices is generally correct}.
Instead, the correct choice is inherently nonlocal, and depends
upon the electrostatic configuration of the entire system.
For example, for configurations in which $\bf P(r)$
is essentially longitudinal, the correct choice is
${\bf P}_{\rm eff}=\bf P$; but if essentially transverse,
then ${\cal E}_{\rm xc}=0$; and for more complicated geometries,
neither simple choice is correct.  This ultra-nonlocality
of the XC potential appears to be an inherent complicating
feature of the exact DFT theory.

I begin by establishing some notation and reviewing some basic
results of Refs.\ \cite{ggg1,ggg2,mo}.  Consider a periodic
insulating crystal with fixed lattice vectors specifying the
unit cell, and an external electron potential
consisting of a periodic and a linear part,
$V({\bf r})=\widetilde{V}({\bf r})-e{\cal E}\cdot\bf r$.
A tilde, as on $\widetilde V$, will be used to indicate a quantity
having the periodicity of the unit cell, and $\cal E$ is a uniform
electric field.  As long as $\cal E$ is not too large, one can with
very good precision identify a physical (although, strictly,
metastable) state of the system, having periodic density, that is
connected to the ${\cal E}=0$ ground state by slow adiabatic
switching of $\cal E$ \cite{mo,nenciu,explan-time}.  Letting
$\widetilde n$ be this periodic density, we can then search for the
{\it non-interacting} KS system for which the effective potential
has the {\it same} linear part (same field $\cal E$) but periodic
part $\widetilde V_{\rm eff}^{\rm KS}$.  These relations are those
of the conventional KS theory applied naively to the periodic
system, and can be summarized as
\begin{equation}
\{\widetilde V,{\cal E}\}\;
\matrix{ \phantom{x} \cr \longleftrightarrow \cr {\rm I} \cr }
\{\widetilde n,{\cal E}\}\;
\matrix{ \phantom{x} \cr \longleftrightarrow \cr {\rm NI} \cr }
\{\widetilde V_{\rm eff}^{\rm KS},{\cal E}\}
\label{eq1}
\end{equation}
(here `I' and `NI' indicate `interacting' and `noninteracting'
respectively).  Alternatively, one can
identify the electronic polarization ${\bf P}$ of the true
interacting system and search for the non-interacting system
that correctly reproduces both $\widetilde n$ and $\bf P$;
that is,
\begin{equation}
\{\widetilde V,{\cal E}\}\;
\matrix{ \phantom{x} \cr \longleftrightarrow \cr {\rm I} \cr }
\{\widetilde n,{\bf P}\}\;
\matrix{ \phantom{x} \cr \longleftrightarrow \cr {\rm NI} \cr }
\{\widetilde V_{\rm eff}^{\rm GGG},{\cal E}_{\rm eff}\}\;\;.
\label{eq2}
\end{equation}
This is the approach of GGG \cite{ggg1,ggg2}.
Unless dictated by symmetry, as for a centrosymmetric crystal in
zero field, there is no reason to expect $\cal E_{\rm eff}=E$, any
more than we expect some particular Fourier component of $\widetilde
V_{\rm eff}$ to match that of $\widetilde V$.  Thus, the
``exchange-correlation field'' defined as ${\cal E_{\rm xc}
=E_{\rm eff}-E}$ is generally non-zero.
(It is understood that ${\cal E}_{\rm xc}$
is not a true electric field, since it acts only on the electrons.)
In either case, Eq.\ (1) or (2), it has to be supposed that the
non-interacting system is also an insulator \cite{explan-ge}, and
that the field $\cal E$ or $\cal E_{\rm eff}$ acting on this
non-interacting insulator is again small enough so that a
metastable state is well defined \cite{explan-time}.

\begin{figure}
\epsfxsize=3.3 truein
\centerline{\epsfbox{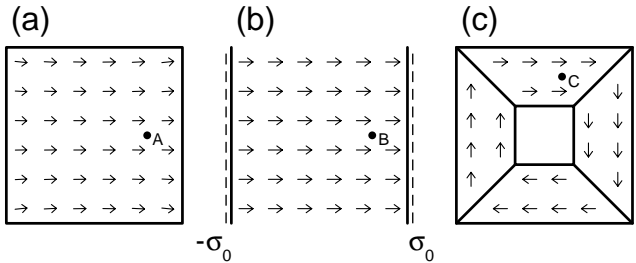}}
\caption{Sample geometries of a spontaneously polarized material.
Details are given in text.  (a) Cubic sample with spontaneous
polarization along $+\hat x$, partially reduced by depolarization
fields arising from surface charges.  (b) Slab geometry in which
surface charges ${\bf P}_{\rm tot}\cdot\hat n$ are precisely
cancelled by appropriately chosen external planar charges
$\mp\sigma_0$.  (c) Geometry composed of four domains in such a
way that $\nabla\cdot{\bf P}=0$.}
\label{fig1}
\end{figure}

Either procedure, Eq.\ (1) or (2), is perfectly sensible in the
abstract, but we now have to investigate whether and how it
might apply to the case of a more realistic non-uniform
configuration of a dielectric material.  Consider, for example,
the three geometries sketched in Fig.\ 1; we consider first the
electrostatic configuration of the physical (interacting) system
for each case.  Figure 1(a) shows
a cubic sample of a spontaneously polarized material.  For
definiteness, let us take this material to be BaTiO$_3$ in the
cubic perovskite structure with the atomic coordinates frozen
as follows: the unit cell is ideal cubic, the Ba
nuclei lie at the cube corners, the O nuclei lie on the cube
face centers, and the Ti nuclei are displaced by a constant
distance $\delta_{\rm Ti}=0.05\,$\AA\ along $\hat x$ from the
cube centers.  This material retains a gap of several eV and
has a spontaneous (zero-field) polarization
${\bf P}_{\rm tot}^{(0)}={\bf P}_{\rm ion}+{\bf P}^{(0)}$ with
both components ${\bf P}_{\rm ion}$ (nuclear plus core) and
${\bf P}^{(0)}$ (valence electronic) lying along $+\hat x$ \cite{zhong}.
The geometry of Fig.\ 1(a) is such that the surface discontinuities
in $\bf P_{\rm tot}(r)$ give rise to macroscopic surface charges on the
left and right faces \cite{explan-common-gap}, generating an
electric field that partially depolarizes the sample.  Thus,
the electronic polarization at the point A in Fig.\ 1(a) will
be somewhat reduced from $P^{(0)}$.  In the configuration of
Fig.\ 1(b), this is avoided by placing infinitely thin sheets of
additional external charge $\mp \sigma_0=\pm P_{\rm tot}^{(0)}$ on
the left and right faces respectively, precisely canceling the
depolarization field \cite{explan-1b-yz}.  Thus, the macroscopic
electric field vanishes at point B, and the electronic polarization at this
point is just ${\bf P}=P^{(0)}\hat x$.  Finally, Fig.\ 1(c) shows a
different configuration in which the internal fields also vanish.
This time the sample is comprised of four domains in which the
displacements of the Ti nuclei are along $\hat e({\bf r})$ with
$\hat e=+\hat x$, $-\hat y$, $-\hat x$, and $+\hat y$ in the top,
right, bottom, and left domains, respectively.  Clearly the solution
is ${\cal E}({\bf r})=0$ since then ${\bf P(r)}=P^{(0)}\hat e(\bf r)$
and thus $\nabla\cdot{\bf P}_{\rm tot}=
\nabla\cdot{(\bf P}_{\rm ion}+{\bf P})=0$, which is consistent with
${\cal E}=0$.  {\it Thus, the local
conditions at point B of Fig.\ 1(b) and point C of Fig.\ 1(c) are
identical: ${\cal E}=0$ and ${\bf P}=P^{(0)}\hat x$.}

Of course it must be assumed that the samples in Fig.~1 are
sufficiently large that macroscopic fields can be defined.  So,
when we speak of ``point A,'' we really refer to a region,
large compared to atomic dimensions but small compared to
sample dimensions, in which a periodic $\widetilde V$ and field
$\cal E$ can be identified.  But note, also, that for a macroscopic
sample having the configuration of Fig.\ 1(a), the electrostatic
potential difference between the left and right faces may greatly
exceed the band gap, so that in principle the ground state would
become metallic.  However, in the context of dielectric theory one
is again much more interested in the metastable \cite{explan-time}
insulating state obtained by starting from a configuration without
macroscopic electric fields, such as that of Fig.\ 1(b), and then
adiabatically restoring the fields.  Throughout this Letter it will
always be assumed that both the physical and the
fictitious KS systems are in such metastable states
\cite{explan-surf-ins}.

Let us now deduce what must be the behavior of $V_{\rm xc}(\bf r)$
for each of the configurations of Fig.\ 1.  The dot labeled `A' in
Fig.\ 2 represents the values of the physical electric field $\cal E$
and polarization $P$ of the interacting system of Fig.\ 1(a) at
point A.  These determine the density $\widetilde n$ at point
A, which must be reproduced by the fictitious KS system at point A.
The dashed curve indicates the locus of values
$({\cal E}_{\rm eff},P_{\rm eff})$ of the KS system that are
consistent with each other and with this given $\widetilde n$.
The choice of Eq.\ (1) corresponds to point A$'$ (insisting that
$\cal E_{\rm eff}=E$), while that of Eq.\ (2) corresponds to
point A$''$ (insisting that $P_{\rm eff}=P$).  Any point on
the dashed curve generates the correct periodic density at A, and
so is a candidate for the state of the KS system at A.

\begin{figure}
\epsfxsize=2.4 truein
\centerline{\epsfbox{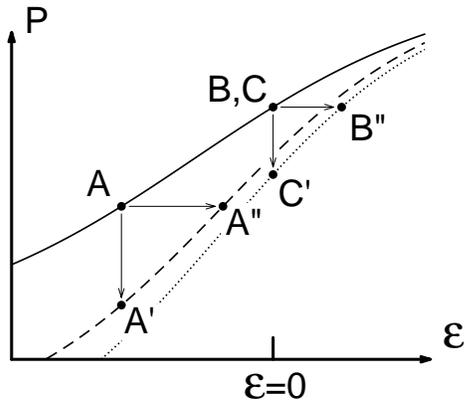}}
\caption{A, B, and C indicate the local state
of the physical (interacting) system at corresponding points
of Fig.\ 1; solid line is the dielectric relation for the
physical material.  A$'$ and C$'$ indicate states of the
Kohn-Sham system having the correct density and field, Eq.\ (1);
A$''$ and B$''$ indicate KS states having the correct density
and polarization, Eq.\ (2).  Dashed and dotted curves indicate
the KS dielectric relations defined under the constraint that
the periodic density match that of the physical system at A,
or B and C, respectively.}
\label{fig2}
\end{figure}

Now comes the crucial point of the argument.
In order to decide which point on this curve should be selected,
it is necessary to inspect the configuration of the system
{\it as a whole}.  This is illustrated by the geometries of
Fig.\ 1(b) and (c), for each of which the correct choice is
easily deduced.  As remarked above, in both cases the physical
system is locally the same (point labeled `B,C').  For the
geometry of Fig.\ 1(b), the translational
symmetry\cite{explan-1b-yz} along $\hat y$ and $\hat z$ insures
that both $\bf P$ and ${\bf P}_{\rm eff}$ lie along $\hat x$.
But in either the physical (interacting) or the KS systems, the
value of $\bf P$ is necessarily related \cite{vks} to the presence
of a macroscopic electronic surface charge ${\bf P}\cdot\hat n$;
and since this electronic charge must be given
correctly in the KS theory, it is safe to conclude that
${\bf P}_{\rm eff}=\bf P$.  Thus, for every point in the
interior of Fig.\ 1(b), the state of the KS system is given
by B$''$ in Fig.\ 2.  On the other hand, in Fig.\ 1(c)
both the physical and KS electric fields must vanish everywhere.
Heuristically, this is suggested by the invariance of
$V_{\rm eff}(\bf r)$ under fourfold rotation.  More precisely, we
can check that a consistent solution is obtained when, for every
point in the interior of Fig.\ 1(c), the state of the KS system is
given by C$'$ in Fig.\ 2.  In this case the polarization {\it is
given incorrectly by the KS theory} at every interior point.
Nevertheless, since $\nabla\cdot{\bf P}_{\rm eff}=0$ everywhere,
such an error is still consistent with the KS system having exactly
the correct $n({\bf r})$.

The example of Fig.\ 1(b) and (c) demonstrates that a knowledge
of the local state of the physical interacting system (i.e.,
a knowledge of the material, and of $\cal E$ or equivalently
of $\bf P$) is {\it inherently insufficient} to determine
the corresponding local state of the KS system.  Thus,
a field ${\cal E}_{\rm xc}$ exists at point B, but not at point C,
in spite of the fact that the local states of the physical
systems at B and C are identical.

In general, the correct correspondence to the KS system must be
determined by an analysis of the electrostatic configuration
as a whole, and need not reduce to either simple choice
(${\cal E}_{\rm eff}=\cal E$ or ${\bf P}_{\rm eff}=\bf P$).
For example, for point A of Fig.\ 1(a), the correct choice
of KS system might correspond to any of the points on the dashed
curve of Fig.~2.  To identify the correct point, an analysis of
the following type would need to be carried through.
Assuming that the physical $\cal E(\bf r)$ and $\bf P(r)$ are
known: (i) At each point in space, obtain $\widetilde n$ from
Eq.\ (1) or (2), and also obtain the dielectric relation
${\bf P}_{\rm eff}={\cal P}_{\tilde n}(\cal E_{\rm eff})$
for a non-interacting system constrained to have fixed
$\widetilde n$ as $\cal E_{\rm eff}$ is varied (e.g.,
the dashed or dotted curve in Fig.\ 2). (ii) Taking first
${\cal E}_{\rm xc}=0$ (i.e., ${\cal E}_{\rm eff}=\cal E$)
everywhere, compute the macroscopic charge density error
$\delta n({\bf r})=\nabla\cdot[\bf P_{\rm eff}(r)-P(r)]$.
(iii) Find a macroscopic field $\cal E_{\rm xc}(\bf r)$
such that, when acting on the system of nonlinear dielectric
materials specified by dielectric responses
${\cal P}_{\tilde n}$, it induces precisely a cancelling
charge density $-\delta n(\bf r)$.

The above is rarely a simple procedure in practice, but
it illustrates the conditions under which a field $\cal E_{\rm xc}$
is needed.  In Fig.\ 1(a) and (b), one finds $\delta n\ne0$,
so the XC field must be present; while for Fig.\ 1(c),
$\delta n=0$ and so ${\cal E}_{\rm xc}=0$.  In fact, it is evident
that for a purely {\it longitudinal} configuration such as
Fig.\ 1(b), Eq.\ (2) applies and the KS field $\cal E_{\rm eff}$ fails
to match $\cal E$; while for a purely {\it transverse}
configuration as in Fig.\ 1(c), Eq.\ (1) applies and instead the KS
polarization $\bf P_{\rm eff}$ fails to match $\bf P$.  It is easy
to see that infrared-active longitudinal and transverse phonons in
a polar insulator behave in a way analogous to Figs.\ 1(b) and (c),
respectively.

A few comments may be in order.
(i) According to the theory of Refs.~\cite{ksv,vks,resta1,om},
the polarization $\bf P$ is actually only well-defined
modulo $e{\bf R}/\Omega$, where $\bf R$ is a lattice vector and
$\Omega$ is the cell volume.  For the puposes of the arguments
given here, we assume that the polarization differences [e.g.,
between domains in Fig.~1(c)] are sufficiently small that there
is no difficulty in choosing the correct branch of $\bf P$.
(ii) When constructing the KS potential for the configurations
of Fig.~1, attention would also have to be given to the microscopic
details of the choice of $V_{\rm xc}(\bf r)$ at surfaces and
interfaces.  However, {\it local} modifications to $V_{\rm xc}(\bf r)$
at a surface or interface will not affect the macroscopic electronic
charge appearing there \cite{vks}, so that the arguments about
macroscopic fields will not be affected.
(iii) In all of the arguments given here, the lattice coordinates have
remained clamped.  A realistic theory of dielectric
materials would of course have to take into account the
lattice relaxation effects.

In summary, it has been shown that the exact KS potential has an
ultra-nonlocal dependence upon the electronic charge density,
and this dependence does not take such a form that it can be
automatically simplified by reference to the local electronic polarization.
In some respects, the present conclusions are disappointing.
One potentially appealing motivation for introducing \cite{ggg1,ggg2}
the expanded HK correspondence
$\{\widetilde V,{\cal E}\}\leftrightarrow \{\widetilde n,{\bf P}\}$,
expressed explicitly in Ref.\ \cite{mo}, is a hope that the generalized
XC functional of density and polarization might be more localized
than the conventional functional of density alone.  The present
conclusions unfortunately do not support this view.  In fact, they
reinforce previous indications that the true KS functional $V_{\rm
xc}[n]$ is, in general, an extraordinarily non-local and irregular
functional of the density \cite{mazin}.  As for the peculiar properties of
$V_{\rm xc}[n]$ demonstrated here, there seems to be little hope of
capturing these in practical approximations, even semilocal ones
such as the weighted- or average-density approximations
\cite{AWDA}.  And indeed, it is not obvious that it would be
desirable to do so.  Of course, the approximate local and
generalized-gradient functionals currently in use do not have these
peculiarities.

This work was supported by NSF Grant DMR-96-13648.
I wish to thank R.M.~Martin for a conversation that stimulated
the present work.


\begin{references}

\bibitem{ksv} R.D.~King-Smith and D.~Vanderbilt,
Phys. Rev. B {\bf 47}, 1651 (1993).

\bibitem{vks} D.~Vanderbilt and R.D.~King-Smith,
Phys. Rev. B {\bf 48}, 4442 (1993).

\bibitem{resta1} R.~Resta, Rev. Mod. Phys. {\bf 66}, 899 (1994).

\bibitem{om} G.~Ortiz and R.M.~Martin,
Phys. Rev. B {\bf 49}, 14202 (1994).

\bibitem{hk} W.~Kohn and L.J.~Sham,
Phys. Rev. {\bf 140}, A1133 (1965).

\bibitem{ks} P. Hohenberg and W.~Kohn,
Phys. Rev. {\bf 136}, B864 (1964).

\bibitem{ggg1} X.~Gonze, Ph.~Ghosez and R.W.~Godby,
Phys. Rev. Lett. {\bf 74}, 4035 (1995).

\bibitem{ggg2} X.~Gonze, Ph.~Ghosez and R.W.~Godby,
Phys. Rev. Lett. {\bf 78}, 294 (1997).

\bibitem{aulbur} W.G.~Aulbur, L.~J\"onsson, and J.W.~Wilkins,
Phys. Rev. B {\bf 54}, 8540 (1996).

\bibitem{resta2} R.~Resta, Phys. Rev. Lett. {\bf 77}, 2265 (1996).

\bibitem{mazin} I.I.~Mazin and R.E.~Cohen, Ferroelectrics
{\bf 194}, 263 (1997).

\bibitem{mo} R.M.~Martin and G.~Ortiz,
Phys. Rev. B, in press.

\bibitem{nenciu} G.\ Nenciu, Rev. Mod. Phys. {\bf 63}, 91 (1991).

\bibitem{explan-time} We shall say that a system in an electric field
is meta\-stable if the characteristic time for tunneling of electrons
from valence to conduction states exceeds some long time (e.g.,
10$^3$ yr).

\bibitem{explan-ge} As is well known, there are exceptions such as
Ge, for which the KS system is metallic [see, e.g., R.W.~Godby and
R.J.~Needs, Phys. Rev. Lett. {\bf 62}, 1169 (1989)].  Such systems
have to be excluded from the considerations given here.

\bibitem{zhong} R.D.~King-Smith and D.~Vanderbilt,
Phys. Rev. B {\bf 49}, 5828 (1994);
W.~Zhong, R.D.~King-Smith, and D.~Vanderbilt,
Phys. Rev. Lett. {\bf 72}, 3618 (1994).

\bibitem{explan-common-gap} It is also assumed here that the surfaces
are insulating, so that the surface charge is related to the subsurface
polarization in the obvious way.
See also Ref.\ \cite{vks}.

\bibitem{explan-1b-yz} Fig.\ 1(b) is intended to suggest that the
system extends indefinitely in the $\hat y$ and $\hat z$ directions.
To be more conservative, we can let the system have large but finite
dimensions in these directions.

\bibitem{explan-surf-ins} Strictly speaking, it is also necessary to
assume that the surfaces of the sample remain insulating.

\bibitem{AWDA} O.~Gunnarsson, M.~Jonson, and B.I.~Lundqvist,
Phys. Rev. B {\bf 20}, 3136 (1979).

\end{references}
\end{document}